\documentclass[aps,prl,reprint,showpacs,superscriptaddress]{revtex4-1}
\usepackage{units}
\usepackage{amsmath}
\usepackage{amssymb}
\usepackage{graphicx}
\usepackage{textcomp}
\usepackage{color}

\begin{document}

\title{Universal sign-control of coupling in tight-binding lattices}

\author{Robert Keil}
\email{robert.keil@uibk.ac.at}
\affiliation{Institut f\"ur Experimentalphysik, Universit\"at Innsbruck, Technikerstrasse 25, 6020 Innsbruck, Austria}
\author{Charles Poli}
\affiliation{Department of Physics, Lancaster University, Lancaster LA1 4YB, United Kingdom}
\author{Matthias Heinrich}
\affiliation{Institute of Applied Physics, Abbe Center of Photonics, Friedrich-Schiller-Universit\"{a}t Jena, Max-Wien-Platz 1, 07743 Jena, Germany}
\author{Jake Arkinstall}
\affiliation{Department of Physics, Lancaster University, Lancaster LA1 4YB, United Kingdom}
\author{Gregor Weihs}
\affiliation{Institut f\"ur Experimentalphysik, Universit\"at Innsbruck, Technikerstrasse 25, 6020 Innsbruck, Austria}
\author{Henning Schomerus}
\affiliation{Department of Physics, Lancaster University, Lancaster LA1 4YB, United Kingdom}
\author{Alexander Szameit}
\affiliation{Institute of Applied Physics, Abbe Center of Photonics, Friedrich-Schiller-Universit\"{a}t Jena, Max-Wien-Platz 1, 07743 Jena, Germany}

\date{\today}

\begin{abstract}
We present a method of locally inverting the sign of the coupling term in tight-binding systems, by means of inserting a judiciously designed ancillary site and eigenmode matching of the resulting vertex triplet. Our technique can be universally applied to all lattice configurations, as long as the individual sites can be detuned. We experimentally verify this method in laser-written photonic lattices and confirm both the magnitude and the sign of the coupling by interferometric measurements. Based on these findings, we demonstrate how such universal sign-flipped coupling links can be embedded into extended lattice structures to impose a $\mathbb{Z}_2$-gauge transformation. This opens a new avenue for investigations on topological effects arising from magnetic fields with aperiodic flux patterns or in disordered systems.
\end{abstract}


\newcommand\bfig{\begin{figure}}
\newcommand\efig{\end{figure}}

\maketitle

Ever since the landmark conjecture of Aharanov and Bohm it is well known that charged particles on closed loops around a magnetic field acquire a phase governed by the ratio of the encircled flux to one flux quantum \cite{AharonovBohm:ABEffect}. 
In two-dimensional tight-binding lattices, this round-trip phase is distributed over the individual hopping links in a manner determined by the choice of gauge \cite{Hofstadter:HofstadterButterfly}. If only real-valued hopping coefficients are permitted, the gauge falls into the $\mathbb{Z}_2$-group and produces phases in multiples of $\pi$, i.e., plaquettes pierced by multiples of half a flux quantum, which can lead to particularly rich physics. For instance, square lattices with homogeneous $\pi$-phase fluxes and dimerised hopping amplitudes have been predicted to host fractional charges with fractional exchange statistics \cite{Seradjeh:ChargeFractionalisationsPiFluxSquareLattice,Seradjeh:FractionalExchangePiFluxSquareLattice}. Under the influence of disorder, such $\pi$-flux square lattices exhibit zero energy modes with critical behaviour \cite{Hatsugai:DisorderdPiFluxSquareZeroModes,Fukui:PiFluxSquareDisorderCriticalBehaviour,Motrunich:ParticleHoleSymmetricLocalisation,Mudry:PiFluxSquareWeakDisorder}. In other geometries, $\mathbb{Z}_2$-gauges can be mapped to the physics of interacting lattice spins \cite{Kitaev:AnyonsHoneycomb} or to the exchange phase of identical fermions on a graph \cite{Harrison:QuantumStatisticsonGraphs}. However, a direct implementation of such large magnetic fluxes in solids exceeds current experimental capabilities by orders of magnitude. Therefore, one has to resort to artificial solids, such as atoms in optical potentials \cite{Dalibard:ReviewArtificialGaugeFieldAtoms}, molecules on metal surfaces \cite{Gomes:MolecularGraphene}, arrays of microwave resonators \cite{Bellec:LinearStrainedGraphene} or optical waveguide lattices \cite{Rechtsman:ArtificialMagField,Rechtsman:FloquetTopologicalInsulators}, to mention a few, and employ artificial gauge transformations for an experimental test of the expected dynamics. Notably, the crucial aspect for realising the $\mathbb{Z}_2$-gauge is the capability to change the sign of the hopping or coupling rate in these systems, which is inherited from the underlying physical processes, at will. Control over this sign can be acquired by dynamical modulation of the lattice \cite{Dunlap:DynLocSolidProposal,Longhi:DynLoc,Lignier:DynLocOpticalLattice,Zhang:CoherentDestructionofTunneling_InducedArrays,Struck:SimulatingMagnetismOpticalLattice,Zeuner:MasslessDiracParticles}, Raman-assisted tunneling \cite{Jaksch:MagFieldOptLatticeRamanTransition,Aidelsburger:StagMagFieldOptLattice} or embedding defects in infinite arrays \cite{Zeuner:NegCoupDefects}. The latter method is only effective for one particular configuration, whereas the former two affect the entire lattice in a periodic manner and require dynamic control or additional pump lasers, respectively. 
 
So far no approach has been known, however, which permits changing the sign of coupling in arbitrary geometries without relying on periodicity and, thus, allowing the implementation of \textit{any} $\mathbb{Z}_2$-gauge transformation.
In this letter, we propose and experimentally demonstrate such a method which operates entirely on the level of individual links. A sign-flip of the coupling is achieved via insertion of a single defect site on the targeted link and matching the eigenmodes of the modified system to the ones of the target lattice. The concept is based solely on static adjustments and can be universally applied to any tight-binding system, given that additional sites can be inserted and the on-site potential can be controlled.

We start with a tight-binding Hamiltonian of a two-dimensional lattice governed by the adjacency matrix $\mathbf{T}$:
\begin{equation}
H=-\sum_{\left\langle j,k\right\rangle}T_{jk}\mathrm{e}^{i\theta_{jk}}a_j^{\dagger}a_k+\textrm{H.c.},
\nonumber
\end{equation}
with $a_j$ as the particle annihilation operator on site $j$ and $\left\langle j,k\right\rangle$ denoting ordered pairs. The phases $\theta_{jk}$ account for magnetic fluxes threading the lattice. In particular, for a plaquette $P$ pierced by the flux $\Phi_P$, the sum over all phases around the plaquette must yield $e\Phi_P/c\hbar$ \cite{Hofstadter:HofstadterButterfly,Seradjeh:ChargeFractionalisationsPiFluxSquareLattice}
(see Fig.~\ref{Setup}\textbf{(a)} for an example).
The specific distribution of phases around $P$ can be chosen arbitrarily and constitutes the gauge degree of freedom. If all fluxes in the system are integer or half-integer multiples of a flux quantum, that is $\forall P:\Phi_P=\nu_P hc/e,\nu_P\in\left\{0,\pm\nicefrac{1}{2},\pm 1,\ldots\right\}$, the associated round-trip phases are multiples of $\pi$. Therefore, such a scenario can be gauged to have only symmetric real hoppings of either positive or negative sign, forming the $\mathbb{Z}_2$-group of gauges (Fig.~\ref{Setup}\textbf{(b)}). 

Facilitating the exploration of all possible $\mathbb{Z}_2$-gauge fields requires an independent control over the signs of all hoppings in the lattice. In what follows, we present a method for exactly that purpose.
To be specific, we use the notation of evanescently coupled optical waveguides, where coupling is positive by default, and show how negative coupling can be implemented in such systems. The dynamics of two identical single-mode waveguides coupled with strength $\kappa>0$ (in the frame co-moving with the mode's propagation constant along the spatial coordinate $z$) is governed by coupled-mode equations \cite{Okamoto:Fundamentals}:
\begin{equation}
i\frac{\mathrm{d}a_{1(2)}}{\mathrm{d}z}+\kappa a_{2(1)}=0,
\label{coupledmode}
\end{equation}
with discrete field amplitudes $a_1$ and $a_2$ guided in the two channels. This system is equivalent to two potential wells, each supporting one eigenmode, with tunneling rate $\kappa$ between them and $z$ representing the time axis. Such a coupler features a pair of stationary solutions $\left(a_1(z),a_2(z)\right)^{\intercal}\equiv\mathbf{u_{\pm}^{\kappa}}\exp\left(i\beta_{\pm}^{\kappa}z\right)$ with eigenvalues $\beta_{\pm}^{\kappa}$ and eigenmodes $\mathbf{u_{\pm}^{\kappa}}$. From Eq.~(\ref{coupledmode}) readily follows:
\begin{equation}
\beta_{\pm}^{\kappa}=\pm\kappa;\;\mathbf{u_{\pm}^{\kappa}}=\frac{1}{\sqrt{2}}\left(1,\pm 1\right)^{\intercal},\nonumber
\end{equation}
which is the characteristic pair of symmetric and antisymmetric eigenmodes (see Fig.~\ref{Setup}\textbf{(c)}). For negative coupling $\kappa\rightarrow-\kappa$, the eigenvalues of the pair are exchanged: 
$\beta_{\pm}^{-\kappa}=\beta_{\mp}^{\kappa}$
, as shown in Fig.~\ref{Setup}\textbf{(d)}. 
The aim is now to find a structure consisting entirely of positively coupled units, which has the same eigenmodes as the idealized negative coupler and, therefore, exhibits identical dynamics.

\bfig
\includegraphics[width=\columnwidth]{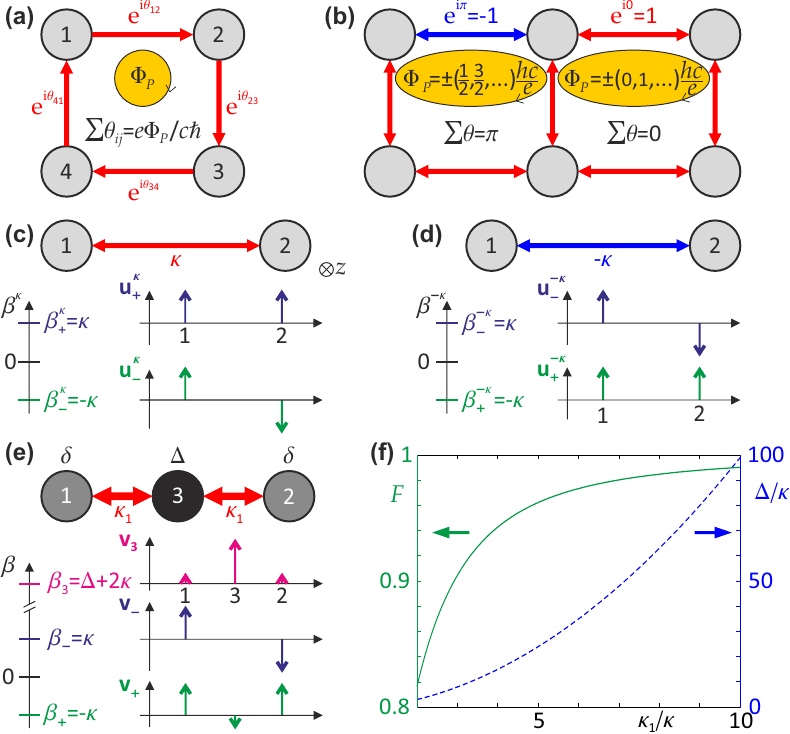}
\caption{\label{Setup}\textbf{(a)} Distribution of Aharanov-Bohm phases on a lattice cell pierced by flux $\Phi_P$. \textbf{(b)} $\mathbb{Z}_2$-gauge: Flux phases of $0$ and $\pi$ can be realised by positive and negative real hoppings. \textbf{(c)} Conventional coupler with identical waveguides and its eigenmodes. Light propagates along the $z$-axis into the image plane. \textbf{(d)} Coupler with negative coupling. \textbf{(e)} Three-site link with detunings $\delta$ and $\Delta$, approximating the negative coupler. Eigenvalues and eigenmodes were calculated for $\delta=\kappa$, $\Delta=\kappa_1^2/\kappa-\kappa$ and $\kappa_1=5\kappa$. \textbf{(f)} Overlap of the symmetric matched eigenmodes of the three-site system and the negative coupler (solid line) and required detuning of the central site (dashed) vs. $\kappa_1$.}
\efig
To this end, we consider an additional waveguide, inserted between the two sites $1$ and $2$ with its propagation constant detuned by $\Delta$. The outer sites may likewise be detuned by some value $\delta$ (see Fig.~\ref{Setup}\textbf{(e)}). The sites are coupled with rate $\kappa_1$, which one can expect to be much larger than $\kappa$ for physical realisations of such a system, due to the exponential dependence of the coupling on the separation between the sites \cite{Somekh:ChannelOpticalWaveguides}. At this point, we will neglect next-nearest-neighbour coupling and discuss its influence later on. The eigenmodes $\mathbf{v}$ of this three-site system are determined by the eigenvalue problem:
\begin{equation}
\left(\begin{array}{ccc}
\delta & 0 & \kappa_1\\
0 & \delta & \kappa_1\\
\kappa_1 & \kappa_1 & \Delta
\end{array}\right)\mathbf{v}
=\beta\mathbf{v}.
\label{eigenvalueproblem}
\end{equation}
\begin{figure*}
\includegraphics[width=\textwidth]{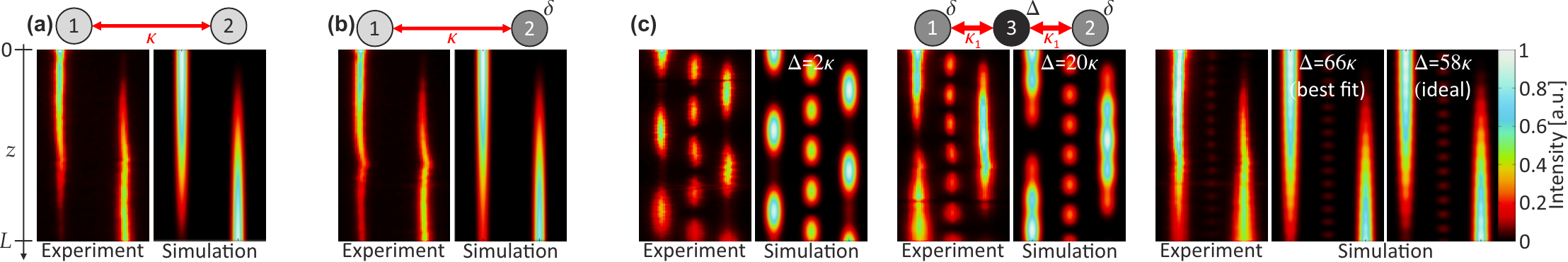}
\caption{\label{Propagation}Light propagation. The measured intensity evolution is compared to the numerical solution of the tight-binding equations. \textbf{(a)} Regular coupler. \textbf{(b)} Coupler detuned by $\delta$. \textbf{(c)} Three-site links with increasing $\Delta$. The other parameters are $\kappa=\unit[0.28]{cm^{-1}}$, $\kappa_1=7.7\kappa$ and $\delta=\kappa$. Each experimental image has been noise filtered and rescaled to its maximum.}
\end{figure*}
One finds that the antisymmetric eigenmode $\mathbf{v_-}=1/\sqrt{2}(1,-1,0)^{\intercal}$ always exists with an eigenvalue of $\beta_-=\delta$, regardless of the other parameters. Thus, setting $\delta=\kappa$ perfectly matches $\mathbf{v_-}$ to the antisymmetric eigenmode of a negative coupler $\mathbf{u_-^{-\kappa}}$ (purple eigenmodes in Figs.~\ref{Setup}\textbf{(d)} and \textbf{(e)}). To acquire a symmetric eigenmode $\mathbf{v_+}$ matching its counterpart for the negative coupler $\mathbf{u_+^{-\kappa}}$, its components must satisfy $v_{+,1}=v_{+,2}$ and its eigenvalue $\beta_+=-\kappa$. Both conditions are met for
\begin{equation}
\delta=\kappa;\;\Delta=\left(\kappa_1^2/\kappa\right)-\kappa.
\label{MatchingCondition}
\end{equation}
In contrast to the antisymmetric mode, however, the matching of the symmetric mode is not perfect, as a part of $\mathbf{v_+}$ resides also on the central site (shown in green in Fig.~\ref{Setup}\textbf{(e)}). 
The fidelity of reproducing the behaviour of a negative coupler can be quantified by the overlap of the normalised symmetric eigenmodes:
\begin{equation}
F\equiv\sum_{j=1}^2v_{+,j}u_{+,j}^{-\kappa}=\left(1+2\left(\kappa/\kappa_1\right)^2\right)^{-\frac{1}{2}}.
\nonumber
\end{equation}
This fidelity grows monotonically with the ratio of the coupling strengths, tending to unity for $\kappa_1/\kappa\rightarrow\infty$, and is plotted as the solid graph in Fig.~\ref{Setup}\textbf{(f)}. Hence, a large ratio $\kappa_1/\kappa$ will be favourable to obtain a high quality of eigenmode matching. In this regime, the third eigenmode $\mathbf{v_3}$ (shown in magenta in Fig.~\ref{Setup}\textbf{(e)}) resides mostly on the defect node, is strongly detuned and, therefore, barely interferes with the other eigenmodes. Note that in practical implementations, the benefit of large ratios $\kappa_1/\kappa$ has to be weighted against the fact that the required detuning $\Delta$ grows quadratically with $\kappa_1$ (dashed graph), for which it becomes increasingly challenging to maintain single-mode guiding in a practical implementation.

In order to demonstrate the viability of our approach, we implemented such systems based on laser-inscribed waveguides in fused silica glass \cite{Miura:LaserWrittenWaveguides}: Pulses from an $\unit[800]{nm}$ mode-locked laser (pulse duration $\tau=\unit[160]{fs}$, repetition rate $\unit[100]{kHz}$, average power $P=\unit[23]{mW}$) were focussed (numerical aperture $0.35$) $\unit[200]{\textrm{\textmu}m}$ into the material, which was moving with velocity $v_0=\unit[110]{mm/min}$, to inscribe $L=\unit[4.9]{cm}$ long waveguides. The first fabricated structure is a conventional positive coupler, with a waveguide separation of $\unit[28]{\textrm{\textmu m}}$. Linearly polarised light from a $\unit[633]{nm}$ laser was butt-coupled into one of its ports and its propagation monitored via fluorescence of color centres \cite{Szameit:Coherence}. For details of the experimental setup see \cite{Szameit:Review_DiscreteOptics} (Figs. 6 and 8 therein). The left panel in Fig.~\ref{Propagation}\textbf{(a)} shows the observed light intensity evolution. Via comparison to the solution of Eq.~(\ref{coupledmode}) (right panel), a coupling strength $\kappa=\unit[0.28]{cm^{-1}}$ can be fitted to the data. In a next step, one waveguide was detuned by reducing its inscription velocity \cite{Bloemer:WaveguideNonlinearity} in order to satisfy the first matching condition $\delta=\kappa$. In such a detuned coupler only $80\%$ of the light power is transmitted to the other site \cite{Okamoto:Fundamentals}. A writing velocity at the detuned site of $v_{\delta}=\unit[104]{mm/min}$ was found to yield this transmission efficacy (see Fig.~\ref{Propagation}\textbf{(b)}). This velocity was subsequently used for all waveguides to be detuned by $\delta$. Finally, the central site was inserted (keeping the distance between the outer sites constant) and its inscription velocity was reduced until the light evolution fitted the pattern expected for eigenmode matching. Fig.~\ref{Propagation}\textbf{(c)} shows the light propagation for three structures with central writing velocities of $v_{\Delta}=\unit[89]{mm/min}$, $\unit[47]{mm/min}$ and $\unit[19]{mm/min}$, respectively. The first one has a rather weak detuning, such that $\kappa_1$ can be unambiguously identified from the oscillation period as $\kappa_1\approx 7.7\kappa$.
For growing $\Delta$ the oscillation between the outer sites is slowed down, whereas the oscillation on the inner site is accelerated, but reduced in amplitude (see second group of images in \textbf{(c)}). The third structure (right group) exhibits a propagation pattern, which facilitates roughly the same rate of light transport between the outer sites as the coupler in \textbf{(a)}. A close inspection reveals that its detuning is actually slightly too large, best fitting to $\Delta_{\mathrm{exp}}\approx 66\kappa$, while the optimal value according to Eq.~(\ref{MatchingCondition}) would be $\Delta_{\mathrm{opt}}\approx 58\kappa$.
However, this deviation from the optimal design parameter does not compromise the functionality as a negative coupler, as only the lower eigenvalue is shifted to $\beta_{+,\mathrm{exp}}\approx -0.77\kappa$, whereas the other eigenvalue $\beta_{-,\mathrm{exp}}=\kappa$ as well as the symmetries of the eigenmodes remain. Therefore, only the rate of coupling is slightly reduced to $(\beta_{+,\mathrm{exp}}-\beta_{-,\mathrm{exp}})/2\approx -0.88\kappa$, showing the robustness of our approach. The measured values for $\kappa_1$ and $\Delta_{\mathrm{exp}}$ translate to a matching fidelity for the symmetric eigenmode of $F_{\mathrm{exp}}\approx 0.987$.

So far, we have only demonstrated that the same magnitude of coupling as in a two-site coupler can be achieved via eigenmode matching in a three-site system. However, no information on the sign of the coupling can be obtained from intensity evolution in the structure alone, as $\kappa$ and $-\kappa$ per definition yield identical patterns.
We therefore turn to an interferometric measurement:
A section of our approximate negative coupler is followed by a conventional positive coupler of the same length, as sketched in the top row of Fig.~\ref{Verification}. 
If the couplings in the two sections have equal magnitudes but opposite signs, their eigenmodes will be swapped at the interface (cf. Fig.~\ref{Setup}). Consequently, the light evolution in the second section will exactly reverse the one in the first and a full revival must occur in the initially excited channel \cite{Longhi:SelfImaging}. 
\bfig
\includegraphics[width=\columnwidth]{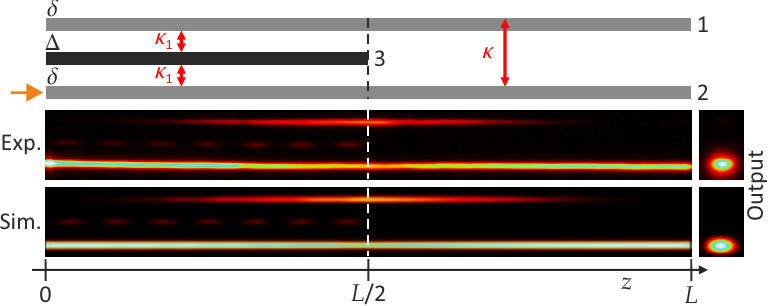}
\caption{\label{Verification}Interferometric phase verification. The three-site structure is superseded by a regular coupler at $z=L/2$. The sign reversal of the coupling leads to refocussing in the injection channel. The parameters are as before, except $\Delta=61\kappa$.}
\efig
To compensate for the slight detuning of the $\mathbf{v_+}$-mode in the previous setting, we implemented the interferometer with $v_{\Delta}=\unit[20]{mm/min}$. The middle row in Fig.~\ref{Verification} shows how the light is imaged back into the input channel during the second half of the propagation. The fidelity of this revival process is quantified by the fraction of power in channel $2$ at the output of the device and is $98.0\%$ in the experiment and $99.92\%$ in the best-fitting numerical simulation (bottom row). Note that if the couplings in both sections had the same sign, the light would continue tunneling from site $2$ to site $1$ and no revival would occur at $z=L$. Therefore, the experimental result shows unequivocally that the coupling in the first part is negative. The very small difference between the theoretical fidelity and $100\%$, expected for ideal negative coupling, is indicative of the high quality of the approximation of the negative coupler by the three-site unit, whereas the deviations in the experiment can be attributed to fabrication inaccuracies breaking the symmetry of the structure. To estimate scaling towards larger systems, one can calculate  the influence of the symmetric eigenmode mismatch (cf. Fig.~\ref{Setup}\textbf{(e)}) on the phase acquired when crossing the negative link. For the parameters of Fig.~\ref{Verification} one obtains a phase error of about $\pi/70$, such that 70 $\pi$-flux cells could be traversed before a gauge error (erasing one flux cell from the system) would occur. These errors can be asymptotically eliminated for systems with $\kappa_1/\kappa\rightarrow\infty$.

We have so far neglected next-nearest-neighbour coupling across the three-site link. This is valid, as evidenced by the observed behaviour: Intuitively, one may expect a coupling rate $\kappa$ between the outer sites, as their distance is the same as in the conventional coupler. Adjusting Eq.~(\ref{eigenvalueproblem}) accordingly would then produce the mode matching conditions $\delta=2\kappa$ and $\Delta=\kappa_1^2/(2\kappa)-\kappa$. Hence, the optimal $\Delta$ would be approximately half the value required by Eq.~(\ref{MatchingCondition}). 
As evident from Figs.~\ref{Propagation}\textbf{(c)} and~\ref{Verification}, the measured propagation dynamics clearly follow Eq.~(\ref{MatchingCondition}) with a high degree of fidelity. Attempting to reproduce the observed patterns with halved $\Delta$ does, on the other hand, not yield any reasonable agreement, which suggests that next-nearest-neighbour coupling is much weaker than $\kappa$ in our system. This observation is fully consistent with another recent experiment, where a suppression of next-nearest-neighbour coupling in a three-site link has been measured directly \cite{Keil:SecondOrderCoupling}.

\bfig
\includegraphics[width=\columnwidth]{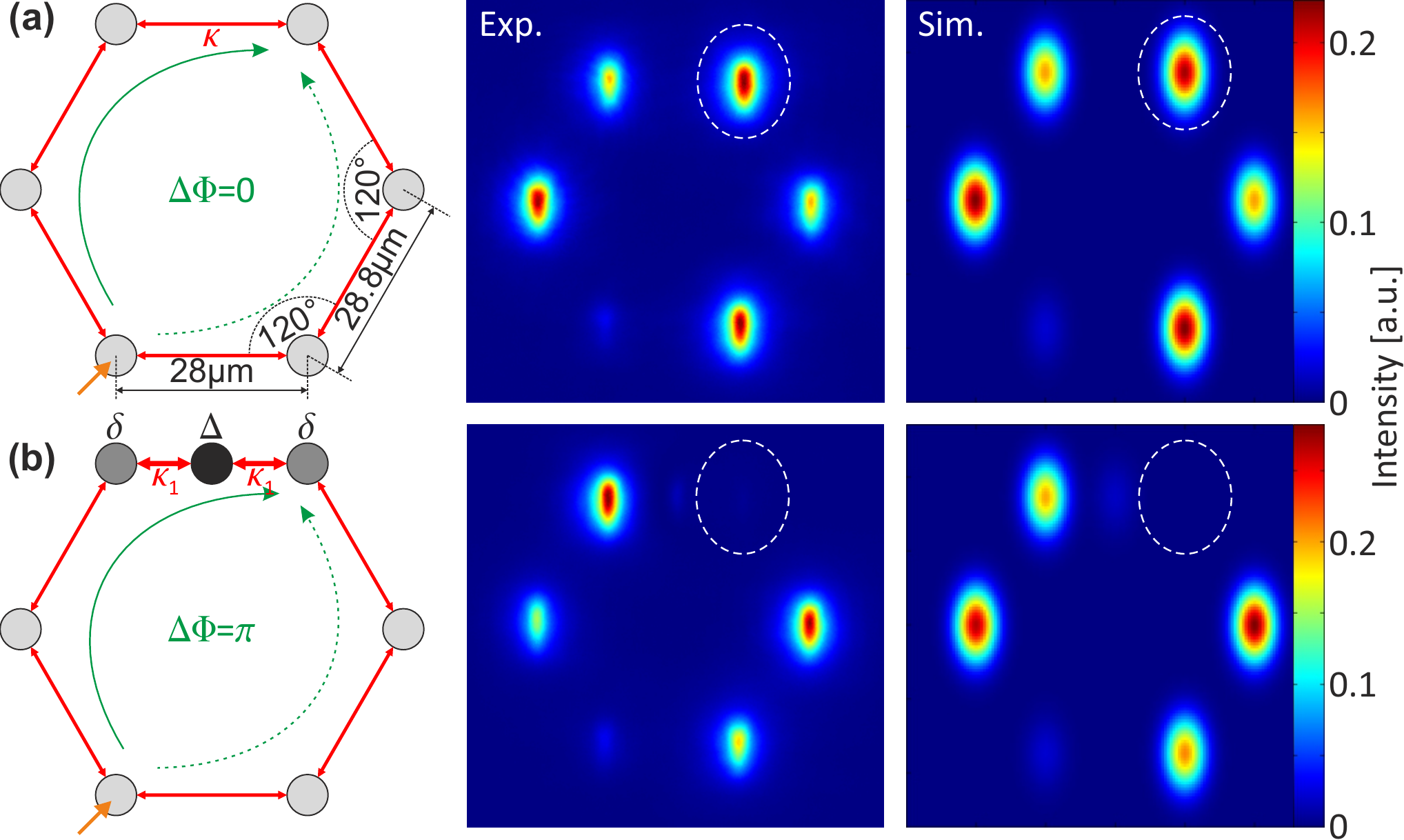}
\caption{\label{Hexagon}Aharanov-Bohm interferometer. \textbf{(a)} Left panel: Hexagon with all-positive couplings. An excitation of the lower left site (orange arrow) leads to constructive interference due to equal phases accumulating on the clockwise (solid arc) and counter-clockwise path (dashed). Central and right panel: Measured and calculated output light distribution, respectively, with simulation parameters $\kappa=\unit[0.27]{cm^{-1}}$ and $\gamma=\unit[0.1]{\textrm{\textmu}m^{-1}}$. \textbf{(b)} Inserting a negative coupling link causes destructive interference. The white ellipses on the output images highlight the difference between the two cases. The defect parameters $\kappa_1\approx 4.1\kappa$ and $\Delta\approx 15\kappa$ are set by $\kappa$ and $\gamma$. Besides replacing the upper link, identical geometries and parameters were used in both systems.}
\efig
Two key advantages of the method of eigenmode matching are that it does not require any assumptions on the surrounding and that the outer sites remain in their positions. Therefore, once the eigenmodes are matched, the approximate negative coupler can be embedded into arbitrary environments to replace a positive coupling link by a negative one. 
We demonstrate this capability in a hexagonal configuration as sketched in the left panel of Fig.~\ref{Hexagon}\textbf{(a)}. From Eq.~(\ref{coupledmode}) follows that positive coupling induces a phase shift of $\pi/2$ when tunneling between adjacent sites, regardless of the direction. Therefore, constructive interference can be expected at the site opposite to the excitation. If, however, one of the links is replaced by an effective negative coupler (Fig.~\ref{Hexagon}\textbf{(b)}), the phase shift across that link will be $-\pi/2$, leading to a phase difference of $\Delta\Phi=\pi$ between the two paths and causing destructive interference. The latter structure imposes a $\mathbb{Z}_2$-gauge transformation corresponding to one half of a magnetic flux quantum passing through the plaquette \cite{Hofstadter:HofstadterButterfly} and constitutes a static photonic Aharanov-Bohm interferometer. The hexagons were implemented in another glass chip with $L=\unit[4.7]{cm}$ and fabrication conditions $\tau=\unit[180]{fs}$, $P=\unit[25]{mW}$ and $v_0=\unit[220]{mm/min}$. The optimal inscription velocities for the defect sites were determined in a procedure analogous to the previous experiments, yielding $v_{\delta}=\unit[206]{mm/min}$ and $v_{\Delta}=\unit[96]{mm/min}$, with the waveguide track of the $\Delta$-site being inscribed twice. The anisotropy of evanescent coupling was compensated by slightly stretching the lattice in the vertical direction, as indicated in Fig.~\ref{Hexagon}\textbf{(a)} \cite{Szameit:Coupling,Poulios:2DQRW}. All other fabrication parameters and experimental settings remained as before. Coupling between non-adjacent sites is taken into account by modelling all couplings with an exponential dependence on inter-waveguide distance $d$: $\kappa(d)\propto\exp\left(-\gamma d\right)$, with decay parameter $\gamma$. As justified above, next-nearest-neighbour coupling across the three-site link is neglected.
The light intensity output from the purely positively coupled hexagon shown in the middle panel of Fig.~\ref{Hexagon}\textbf{(a)} clearly exhibits the expected constructive interference. The parameters $\kappa$ and $\gamma$ are estimated from the best fit to the data, which is shown in the right column. A quite different scenario arises in the hexagon with a negative coupling link, as shown in Fig.~\ref{Hexagon}\textbf{(b)}: Here, virtually no light ($0.4\%$ of the total power in the experiment and $0.05\%$ in the simulation) is detected at the site where the destructive interference takes place. This clearly shows how the presence of the negative coupling link alters the interference condition by inducing a phase shift of $\pi$ on the clockwise path.

The experimental results presented in this work demonstrate how insertion of a defect channel between two tight-binding lattice sites and tailored detuning can be used to reverse the sign of the coupling rate. This approach permits adjusting the sign of individual coupling links in arbitrary tight-binding lattices without the need for dynamic modulation or external control, a key capability for the realisation of arbitrary $\mathbb{Z}_2$-gauge transformations 
in the highly controllable and stable environment of artificial solids. For example, the recently suggested strained Kitaev model with inverted links - corresponding to local flux impurities \cite{Rachel:LandauLevelsMajFermions} - or light trapping via photonic Aharanov-Bohm caging \cite{Longhi:ABCaging} could both be realised via our approach. Negatively coupled links could also be used to endow photonic topological insulators, which can be implemented in a variety of lattice systems \cite{Fang:EffMagFieldDynMod,Rechtsman:FloquetTopologicalInsulators,Hafezi:TopologicalEdgeStatesRingResonators,Schmidt:OptomechanicMagneticFields,Mittal:MeasTopInv}, with local impurity gauge fields. Finally, by introducing interactions, be it via optical nonlinearities in photonic \cite{Lederer:DiscreteSolitonsInOptics} or atomic repulsion in optical lattices \cite{Dalibard:ReviewArtificialGaugeFieldAtoms}, the spectrum of physical phenomena coming into experimental reach will be enriched even further \cite{Weeks:InteractionsPiFluxSquare,Rachel:QuantumParamagnetPiFluxTriLattice}.

\begin{acknowledgments}
We acknowledge support by the European Research Council (ERC, project 257531-EnSeNa), the Canadian Institute for Advanced Research (CIFAR, Quantum Information Science Program), the German Ministry of Education and Research (Center for Innovation Competence program, grant 03Z1HN31), the German Research Foundation (DFG, project SZ276/7-1) and the EPSRC (grants EP/J019585/1 and EP/L01548X/1). R.K. is supported via a Lise-Meitner-Fellowship of the Austrian Science Fund (FWF, project M 1849). M.H. gratefully acknowledges support from the German National Academy of Sciences Leopoldina (grant LPDR 2014-03).
\end{acknowledgments}


%

\end{document}